*Article*

# The Highly Durable Antibacterial Gel-like Coatings for Textiles

**Seyedali Mirmohammadsadeghi [1], Davis Juhas [2], Mikhail Parker [1], Kristina Peranidze [1], Dwight Austin Van Horn [2], Aayushi Sharma [3], Dhruvi Patel [3], Tatyana A. Sysoeva [3], Vladislav Klepov [4], and Vladimir Reukov [1,*]**

[1] Textiles, Merchandising and Interiors, University of Georgia, Athens, GA 30605, USA; Reukov@uga.edu
[2] Chitozan Health, LLC, Summerville, SC 29486, USA; avanhorn@chitozanhealth.com
[3] Department of Biological Sciences, University of Alabama in Huntsville, Huntsville, AL 35758, USA; tatyana.sysoeva@uah.edu
[4] Department of Chemistry, University of Georgia, Athens, GA 30605, USA; klepov@uga.edu
* Correspondence: Reukov@uga.edu

**Abstract:** Hospital-acquired infections are considered a priority for public health systems, which poses a significant burden for society. High-touch surfaces of healthcare centers including textiles provide a suitable environment for pathogenic bacteria to grow, necessitating incorporating effective antibacterial agents into textiles. This paper introduces a highly durable antibacterial gel-like solution, Silver Shell™ finish, which contains chitosan-bound silver chloride microparticles. The study investigates the coating's environmental impact, health risks, and durability during repeated washing. The structure of the Silver Shell™ finish was studied using Transmission Electron Microscopy (TEM) and Energy-Dispersive X-ray Spectroscopy (EDX). TEM images showed a core-shell structure, with chitosan forming a protective shell around groupings of silver microparticles. Field Emission Scanning Electron Microscopy (FESEM) demonstrated the uniform deposition of Silver Shell™ on the surface of fabrics. AATCC Test Method 100 was employed to quantitatively analyze the antibacterial properties of fabrics coated with silver microparticles. Two types of bacteria, *Staphylococcus aureus* (*S. aureus*) and *Escherichia coli (E. coli)* were used in this study. The antibacterial results showed that after 75 wash cycles, a 100% reduction for both *S. aureus* and *E. coli* in the coated samples using crosslinking agents was observed. The coated samples without a crosslinking agent exhibited a 99.88% and 99.81% reduction for *S. aureus* and *E. coli* after 50 washing cycles. AATCC-147 was performed to investigate the coated samples' leaching properties and the crosslinking agent's effect against *S. aureus* and *E. coli*. All coated samples demonstrated remarkable antibacterial efficacy even after 75 wash cycles. The crosslinking agent facilitated durable attachment between the silver microparticles and cotton substrate, minimizing the release of particles from the fabrics. Color measurements were conducted to assess color differences resulting from the coating process. The results indicated fixation values of 44%, 32%, and 28% following 25, 50, and 75 washing cycles, respectively.

## 1. Introduction

Infection diseases continue to be one of the leading causes of death in the world and are a priority in global public health [1, 2]. Based on a recent study, infection-related deaths were the second leading cause of death, accounting for 13.7 million mortalities in 2019. To clarify more, between 10 to 18% of all main causes of death were attributed to 33 bacterial pathogens, including the resistant strains, which can mainly be found in contaminated areas, e.g. clinical wards [3].

Hospital-acquired infections (HAI), as a fourth primary cause of death in the United States, have imposed a considerable burden on healthcare systems in both human resources and economic issues [4]. High-touch surfaces in hospitals and other health centers, such as floors, beds, tables, and textile-based materials, are at risk of being contaminated by pathogenic bacteria. They can easily transfer infectious microbes to patients and hospital workers and cause additional problems [5]. Drug-resistant infections spreading via textile surfaces pose serious threats, and eliminating bacteria from such surfaces has the potential to drastically reduce the spread. To tackle this issue of contaminated textile hospital materials, the common ways are to use disinfectants to remove surface germs and kill bacteria or use disposable items several times a day. Both methods are burdensome for healthcare systems, being time-consuming and less effective than expected. The high transmission rate of pathogens can still infect patients and healthcare workers with weakened immune systems. In addition, antibiotic overuse is increasing, which would cause the emergence of resistant bacteria and various environmental pollutions [6]. Therefore, antibacterial surfaces, particularly in textiles, are becoming increasingly desirable to protect against different types of microorganisms [5, 7-16].

Various hospital textiles such as privacy drapes, upholstery, lab coats, and scrub suits can be an ideal environment for a broad range of microorganisms such as bacteria and fungi to live for months [7]. These textiles are prone to being contaminated by bacteria and transfer them to the human body [8]. Since porosity of these materials creates good conditions for microorganisms to thrive by providing moisture and warmth [5, 8, 9].

There are different types of antibacterial agents that could be used to create antimicrobial textiles. These include chemical compounds, natural extracts, and metal nanoparticles. Chlorinated aromatic structures, N-halamines, and Quaternary Ammonium compounds are the most common chemical compounds for antibacterial fabric finishing in the textile industry [17].

As a chlorinated aromatic structure, one of the chemical antimicrobial agents, Triclosan, has been used in different products such as toothpaste, soaps, shampoos, and facial cleaners. Based on the hydrophobic nature of Triclosan, it has been incorporated into various plastic-based materials, including textiles and air filters. It alters the integrity of the microbe's membranes and disrupts bacterial fatty acid synthesis by inhibiting the activity of enzymes, which are involved in its synthesis. Triclosan's antibacterial properties have been affected by its concentration and the coating process of treated materials. Although it showed higher antibacterial properties when it is concentrated in some studies, the bacteria utilize the multidrug resistance (MDR) efflux pump to expel the drug, thereby reducing its activity. The efficacy of the efflux pump would decrease at lower concentrations, and it would not allow the bacteria to form colonies. However, the main drawbacks of using Triclosan are related to environmental concerns. The presence of Triclosan in the environment can accumulate in aquatic organisms, such as fish and algae. It disrupts the endocrine system of aquatic organisms, affecting their growth and reproduction and increasing bacteria resistance against Triclosan. [18, 19].

N-halamines, such as 2,2,5,5-Tetramethyl-Imidozalidin-4-One (TMIO), create covalent bonds with nitrogen and halogen, offering broad-spectrum antibacterial activity [20, 21]. Quaternary Ammonium Compounds (QAC) disrupt microbial cell membranes through positive-charged nitrogen, exerting biocidal effects against various microorganisms [22, 23]. While these compounds offer effective antibacterial properties, concerns arise regarding environmental impact, potential health risks, and durability during washing or exposure to high temperatures. The selection of antibacterial compounds for fabric finishing requires a careful balance between efficacy, environmental considerations, and durability [17, 24].

Natural substances such as plant extracts, essential oils, and animal products provide various antimicrobial properties. The investigations of the antimicrobial properties of various natural dyes extracted from plants have increased the variety of antimicrobial agents, contributing significantly to the production of antibacterial fabric. Natural plant extracts such as curcumin, basil, clove oil, cyclic oligosaccharides, sericin, and natural plants including onion, aloe vera, and pomegranate have shown unique antimicrobial properties [25-27]. They offer potential benefits, but they have some considerable drawbacks. One key limitation is the variability in composition and potency, influenced by factors like plant growth conditions and extraction methods, leading to inconsistent efficacy. The spectrum of antimicrobial activity may be limited, and concentrations for effectiveness can vary. Degradation over time affects stability and shelf life, while solubility, bioavailability, and formulation challenges may impede practical application.

Using metals and metal salts, such as silver, copper, gold, zinc oxide, and titanium dioxide, for their antimicrobial properties in textiles has a rich historical background. They can be toxic to pathogens even at a very low concentration. Titanium dioxide and zinc oxide are employed for their bactericidal properties, with $TiO_2$ acting as a photocatalyst requiring UV radiation for antibacterial applications. ZnO offers a cost-effective agent with superior whitening and UV-blocking properties on textiles. However, challenges such as the need for UV radiation and potential limitations in killing microbes exist for these coating agents. Copper has excellent antibacterial properties and poses a biocidal effect. While it is not as effective as silver for antibacterial action, copper is still considered a promising material to be incorporated into different substrates, including textiles [17, 28, 29].

Nanoparticles, particularly silver nanoparticles (AgNPs), are prominent for their broad-spectrum antimicrobial activity, self-cleaning properties, and increased dyeability. AgNPs have demonstrated effectiveness against various microbes. Its antibacterial effectiveness is superb compared to other metal-based materials such as Cu-based, ZnO, and $TiO_2$. The minimum bactericidal concentration for silver is lower than other metal-based antibacterial agents, including copper-based ones. It can even have antibacterial efficiency in a small amount [28]. In contrast to antibiotics, which use specific mechanisms of action against pathogens, the biocide effect of silver nanoparticles has been reported with different mechanisms. One mechanism of action is that due to the electrostatic attraction of silver ions to sulfur proteins, they can adhere to the cell wall and cytoplasmic membrane of bacteria, which increases the permeability of the cytoplasmic membrane and disrupts the bacterial envelope [7, 8, 30, 31].

The challenges associated with the production and coating processes of AgNPs include agglomeration, achieving desired morphology and the production of uniform particle size. Agglomeration decreases the stability and efficacy of AgNPs, which needs effective reagents to overcome this challenge [2]. The common reducing agents such as sodium dodecyl sulfate and ascorbate are toxic and hazardous and cause serious environmental problems. Capping with natural compounds has emerged as a solution to promote the stability of AgNPs, particularly when applied to textiles. However, weak interactions on the fabric surface are considered a significant drawback. To address this, the impregnation of capping agents, whether synthetic polymers or natural biological agents, becomes important for enhancing the stability of AgNPs on textiles. Among the various capping agents, chitosan, as a biodegradable and biocompatible biopolymer, has also weak antibacterial and antifungal properties [32-34]. The mechanism of action was reported variably but mainly involves penetrating the phospholipid bilayer of bacteria membranes and disrupting cytoplasmic membrane [35]. Chitosan can also be an ideal reagent, which mitigates agglomeration by introducing electrostatic repulsion and steric hindrance between AgNPs [1]. Moreover, due to having active functional groups in chitosan structure, it forms hydrogen bonds with cotton, hydrolyzed polyester, and nylon fabrics [36]. This multi-functionality of chitosan makes it a promising candidate for achieving stable, well-dispersed, and antimicrobial AgNP-coated textiles, which can tackle the challenges in both production and long-term stability fields [37]. The application of chitosan as a coating agent or stabilizer for silver nanoparticles has been reported in several studies demonstrating better stability and antimicrobial efficacy [2, 10, 35, 38-41]. For instance, Siva et al.

synthesized chitosan-silver nanocomposites with an average particle size of 10 nm to study their antibacterial activity against Staphylococcus *aureus* and *Escherichia coli* [35].

A non-leaching functional textile is generally preferred because it retains its antimicrobial activity for an extended period. Moreover, releasing significant amounts of antimicrobial agents into the environment can be an environmental issue [17]. Therefore, in this study, a gel-like solution containing chitosan-bound silver chloride microparticles (AgMPs) was developed in two distinct average sizes to investigate their impact on antibacterial properties, considering concerns about their environmental impact when they are in nano size. Our focus includes the durability and effectiveness of cotton fabrics coated with silver-chitosan over multiple washes. *Staphylococcus aureus* and *Escherichia coli*, and multidrug-resistant strains (MG and ES), were chosen to study the efficiency of coated fabrics.

## 2. Materials and Methods

### *Materials*

The fabrics used in this study were LS Bleached Cotton Muslin (120 GSM) from Joanne Fabrics. These fabrics were coated with two types of batches. The first is Silver Shell™ batch SS26 obtained from Chitozan Health LLC and produced at a facility in Chesterfield, England. The second one is Silver Shell™ batch AgMP034 obtained from Chitozan Health LLC, Rochester, NY, USA. Fabrics were coated with AgMP034 and SS26, along with a crosslinking agent. To investigate the effect of the crosslinking agent, one sample, coded as A34, utilized the same batch of AgMP034 but without adding a crosslinking agent during a coating process. The synthesis and coating process employed in this study is proprietary of the Chitozan Health LLC, providing silver chloride crystals suspended in a chitosan solution with adjustable sizes from 50 to 350 nm. Tryptic Soy agar was purchased from Merck, Germany. LB broth was purchased from Sigma-Aldrich, USA. AATCC 1993 standard reference detergent was used for washing samples.

### *Characterization of Silver Shell™ Solutions*

The phase identification of thin film samples containing core-shell microparticles was conducted based on powder X-ray diffraction (PXRD) data (Bruker D2 PHASER diffractometer with Cu K$\alpha$ radiation over a 2$\theta$ range 5–65° and step size of 0.02°). Transmission electron microscope JEM100CX-II (JEOL, Ltd., Tokyo, Japan) operated at 100 kV was used to study the core-shell microparticles. Moreover, the element composition was analyzed by Transmission electron microscopy (Hitachi-SU9000EA) equipped with Energy-dispersive X-ray spectroscopy. A drop of a solution was placed on a copper grid and dried overnight at room temperature.

### *Characterization of Coated Fabrics*

*Surface morphology:*

The surface morphology of the uncoated and coated fabrics was observed by field emission scanning electron microscopy (FESEM) using Thermo Fisher Scientific (FEI) Teneo. The Teneo is equipped with energy-dispersive spectroscopy (EDS) that provides elemental analysis. The samples underwent gold coating for improved imaging using the SPI-Module Sputter coater instrument for one minute.

*Color measurements:*

The color coordinates of CIE lab (L*, a*, and b*) for both untreated cotton and Silver Shell™ coated samples were conducted at 5 different places of each one using a Macbeth

Color Eye 7000A Spectrophotometer. L* is related to lightness, a* is a reddish/greenish factor, and b* corresponds to the yellowness/bluish factor [13].

The color strength (K/S) of the samples coated with Silver Shell™ was also measured at all wavelengths from 400 to 700 nm (10nm interval). This value was reported for each sample at 370 nm, the maximum absorbency wavelength. The color strength value was determined through the color matching software's internal computations, employing the Kubelka–Munk equation [42]:

$$\frac{K}{S} = \frac{(1-R)^2}{2R}, \quad (1)$$

where $K$ represents absorption, $S$ denotes scattering, and $R$ stands for reflectance.

*Antimicrobial Fabric Test:*

To analyze the antibacterial properties of fabric silver coating against uncoated controls the AATCC Test Method 100–2012 was performed. *Staphylococcus aureus* (ATCC 6538) and *E. coli* (ATCC 10536) were used as the main test organisms. Uncoated and silver-coated fabric samples were cut as circular swatches, 4.8 cm in diameter. Four swatches were soaked with 1 mL of inoculum, containing approximately $10^6$ CFUs, and placed in sterile 100 mL glass jars. Swatches soaked with 1mL of sterile PBS were used as controls.

Jars were divided into groups: one with zero-time incubation time to quantify the initial number of viable cells, and another with overnight incubation. In the zero-time incubation group, 100 mL of sterile PBS was added immediately after placing bacteria-inoculated swatches. After vigorous shaking, we made serial dilutions of 1/1, 1/10, 1/100, and 1/1000 from 100 μL of the PBS wash of the swatches. 100 μL of each dilution was spread on agar plates in duplicates. Inoculated plates were incubated overnight at 37°C. Formed colonies were counted to estimate densities of viable cells: colony forming units (CFU) per mL of solution (CFU/mL).

A group of jars with overnight incubation was first incubated overnight at 37°C and then underwent the same serial dilution procedure as the zero-time group to establish cell densities (CFU/mL).

The resulting CFUs on agar plates were counted manually and the percentage of reduction (R) was calculated as follows:

$$R\% = \frac{(B-A)}{B} \times 100, \quad (2)$$

where A is the CFUs or colony forming units (CFU/mL) coated fabrics after 24 h and B is the mean of CFUs counted from control samples and coated samples at zero-time contact.

To compare antibacterial properties toward non-pathogenic and pathogenic strains of the same species, we used MG1655 model *E. coli* strain (ATCC 29213) and multidrug-resistant clinical isolate from Sysoeva lab collection (extended-spectrum b-lactamase producing ESBL41 and ESBL146 strains (Lopatkin et al, 2016). Loading of the swatches, incubations, and plating for CFU counting were done analogously to the procedure for other strains as described. Initially, cell cultures of *E. coli* strains in LB were grown for 24 hr reaching an OD of 2.6-3.8 (measured by spectrophotometer Genesys 50) which is equivalent to about 2.6-3.8*$10^8$ CFU/mL. Cells were diluted to bring them to $10^6$ CFU/mL to load $10^6$ per 5 swatches in 1 mL of solution.

To investigate the leaching of fabric silver coating against uncoated controls, we used the AATCC Test Method 147–2016. Using a wire loop, test microorganisms (*S. aureus* and *E. coli*) were streaked on agar plate in 5 streaks spaced approximately 1 cm apart from each other. For each agar plate, the wire loop was submerged only once. As a result, each lower streak contained fewer CFUs than the upper one. Plates were done in duplicates. After that rectangular swatches (25 mm× 50 mm) were pressed on the streak inoculum, and the plates were incubated for 24 h at 37 °C. After incubation, bacterial growth under and on the edges of fabric were investigated.

## 3. Results and Discussion

*Silver Shell™ Solution Characterization*

The sample of Silver Shell™ solution was annealed at 80°C within 2 hours to obtain a film for phase composition study. The PXRD data collected for the material is demonstrated in Figure 1. The diffractogram illustrates the presence of AgCl, $Ag_2O$, and Ag phases. Narrow characteristic peaks corresponding to the identified phases indicate the regions of high crystallinity.

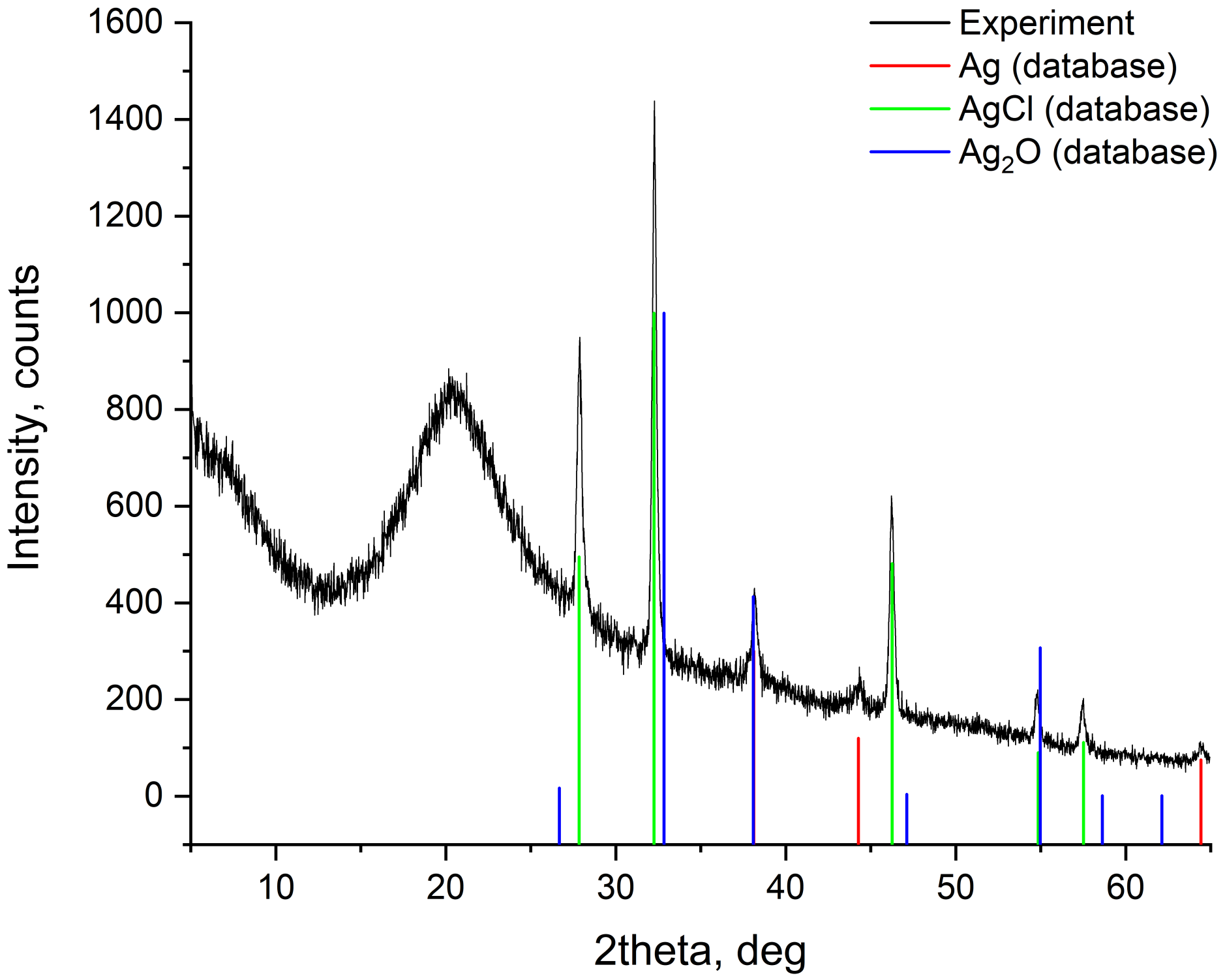

**Figure 1.** PXRD plot for the annealed Silver Shell™ solution.

The TEM images of synthesized SS26 and AgMP034 are shown in Figures 2a, and 2b respectively. According to Figure 2a, the submicron particles are scattered separately and primarily exhibit spherical morphology. However, some multi-shaped particles were also observed, which might be the result of aggregation during the preparation of particles for TEM analysis. The TEM image of AgMP034 (Figure 2b) showed a core-shell structure, in which chitosan formed a protective shell around groupings of microparticles. Based on the TEM images, the median size of particles was found to be 100 nm for SS26 (Figure 2c) and around 202 nm for AgMP034 (Figure 2d). The formation of a chitosan shell is advantageous due to its ability to form hydrogen bonds with the functional groups present in materials like cotton or nylon, thereby facilitating specific interactions and enhancing coating stability.

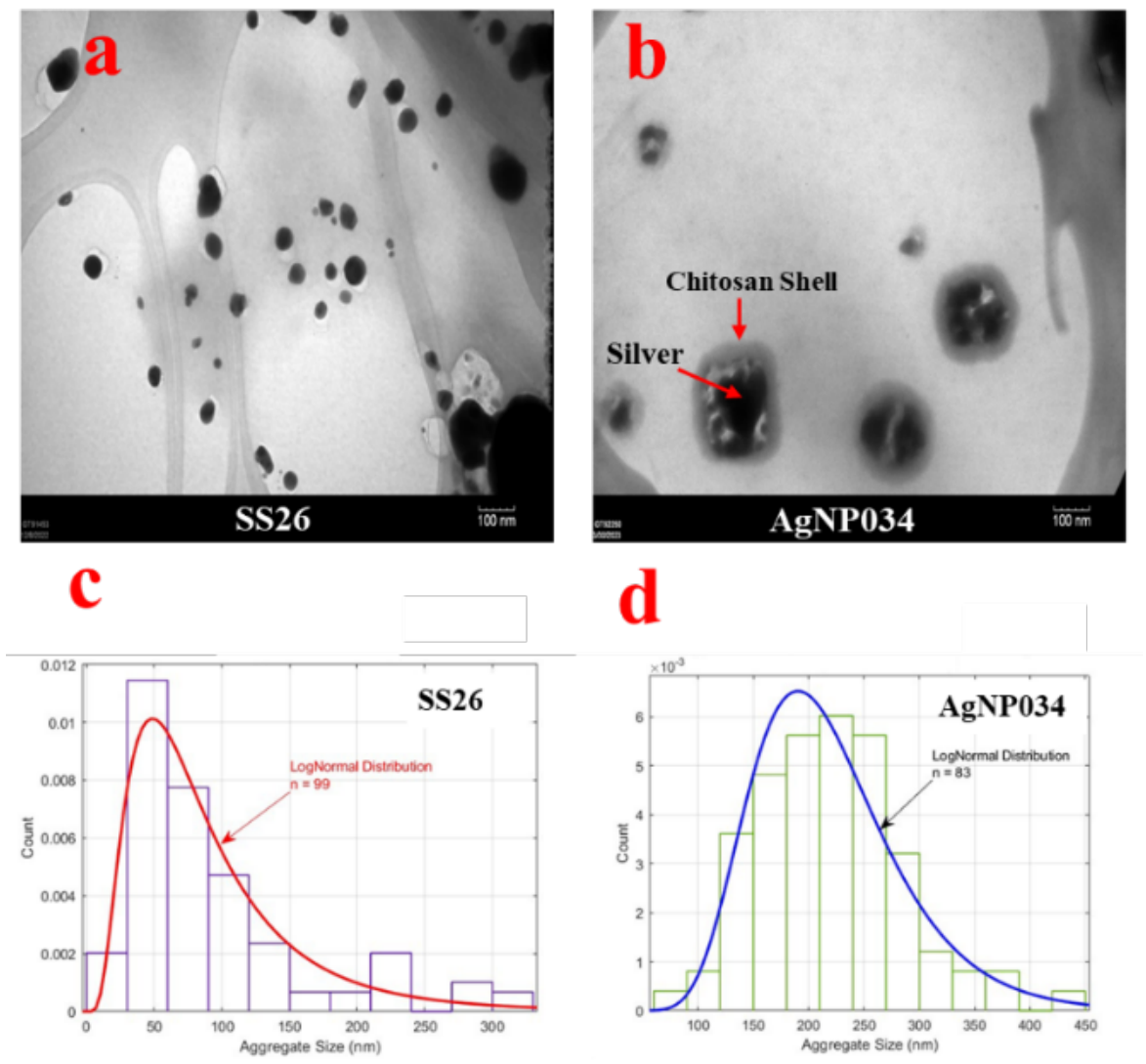

**Figure 2.** TEM images of a) SS26, b) AgMP034; Log Normal Distribution of c) SS26, d) AgMP034.

The analysis of EDX (Figure 3) showed the presence of Ag and Cl elements which are related to the precipitation of AgCl.

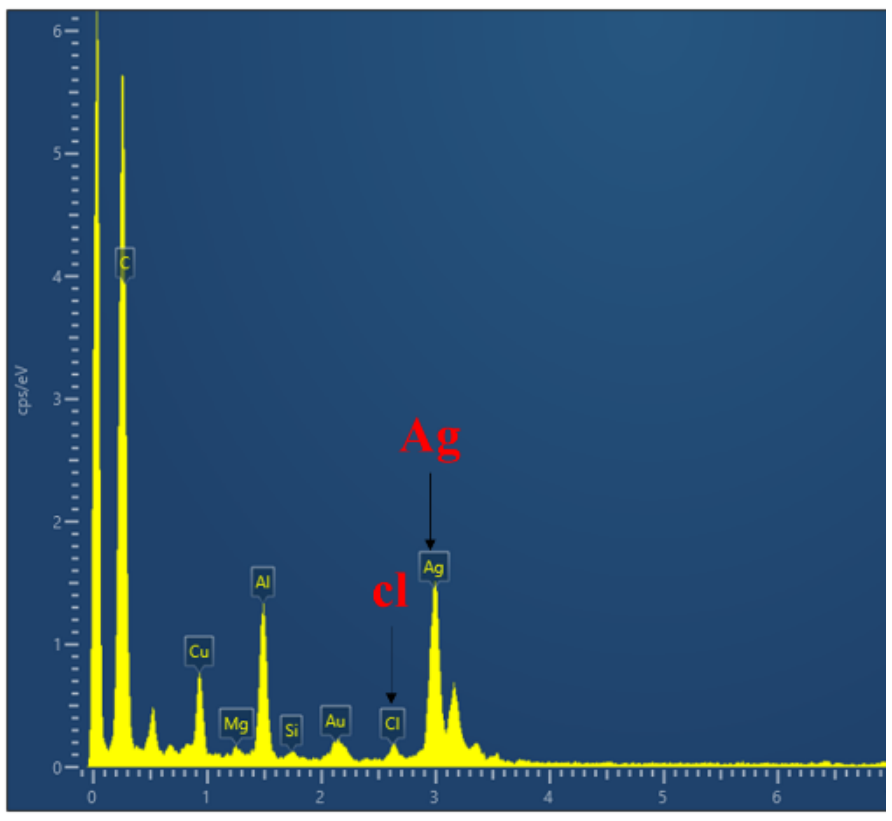

**Figure 3.** EDX analysis of synthesized Silver Shell™ solution.

*Coating Characterization*

Cotton fabrics coated with Silver Shell™ solution were chosen in this work. Generally, cotton and silver particles have physical adsorption [9]. The morphological changes of cotton fabrics coated with Silver Shell™ after 25, 50, and 75 washing cycles were monitored by FE-SEM. As shown in Figure 4a and Figure 5a, Silver Shell™ was uniformly deposited on the cotton fabric surface. The surface of the cotton fabric was relatively smooth after coating. After washing the samples, Figure 4 b-d and Figure 5 b-d, Silver Shell™ began to detach along with the cotton fibers, with the most noticeable detachment observed after 75 washing cycles. However, for SS26 there were some detachments after 50 washing cycles. The results showed that the Silver Shell™ has a strong bond with cotton fabrics due to the presence of hydrogen bonding between chitosan and Cotton.

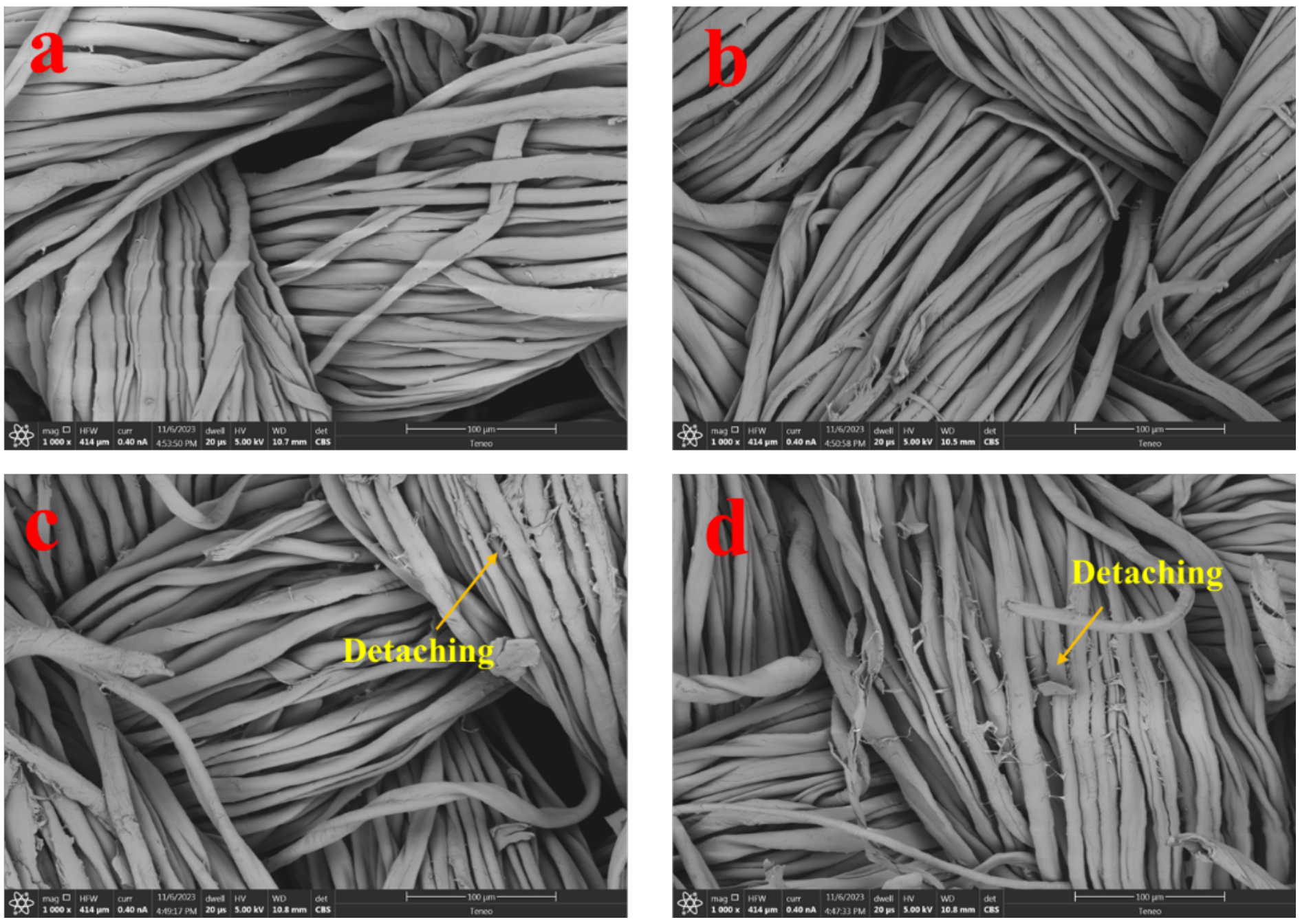

**Figure 4.** SEM images of SS26 with different washing cycles: a) no washing, b) 25 cycles, c) 50 cycles, d) 75 cycles.

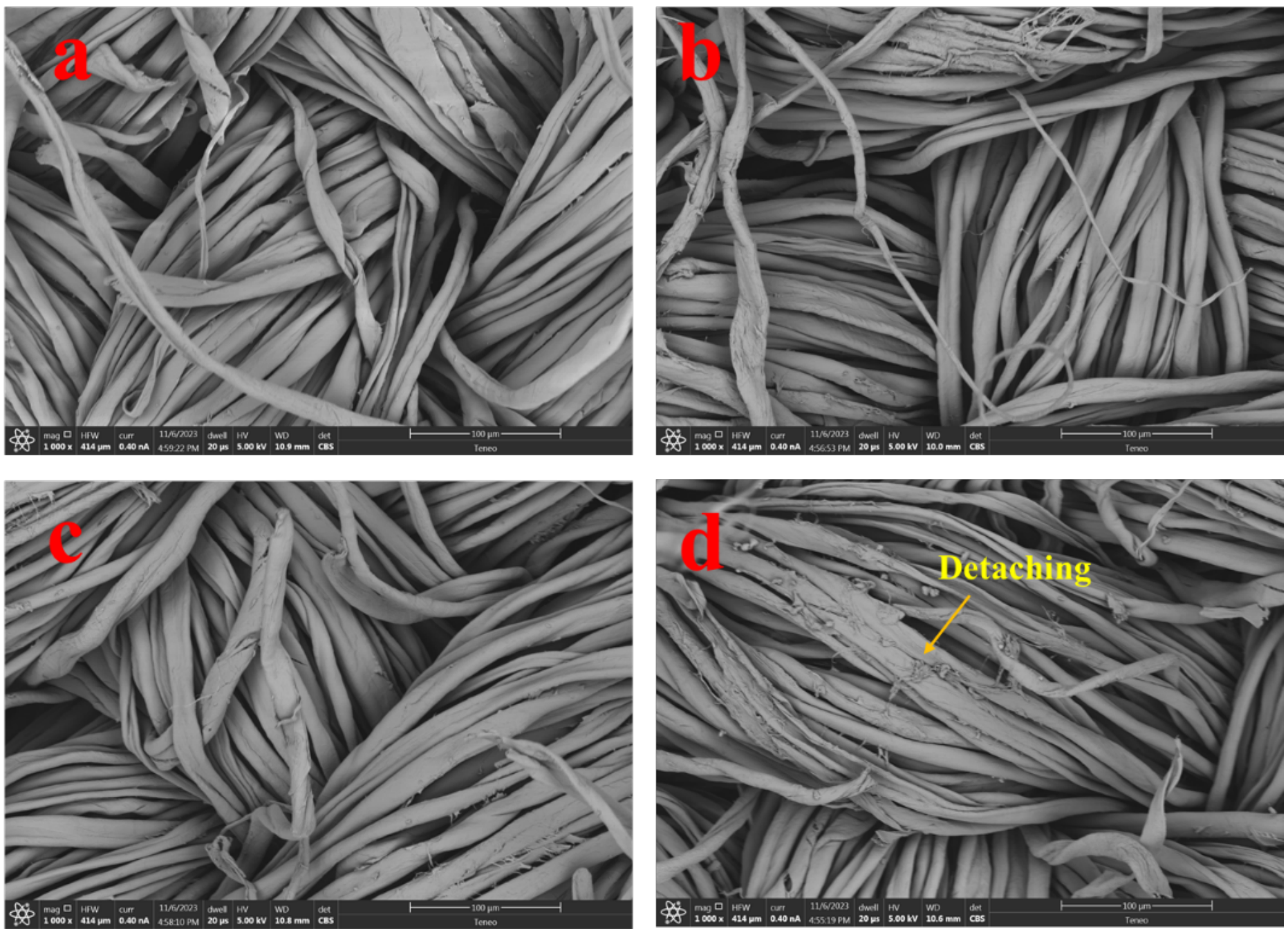

**Figure 5**. SEM images of AgMP034 with different washing cycles: a) no washing, b) 25 cycles, c) 50 cycles, d) 75 cycles.

Figure 6 shows the result of EDX analysis of AgMP034 without washing. The weight % of Ag was 0.2% as the silver coverage on coated fabric was very low.

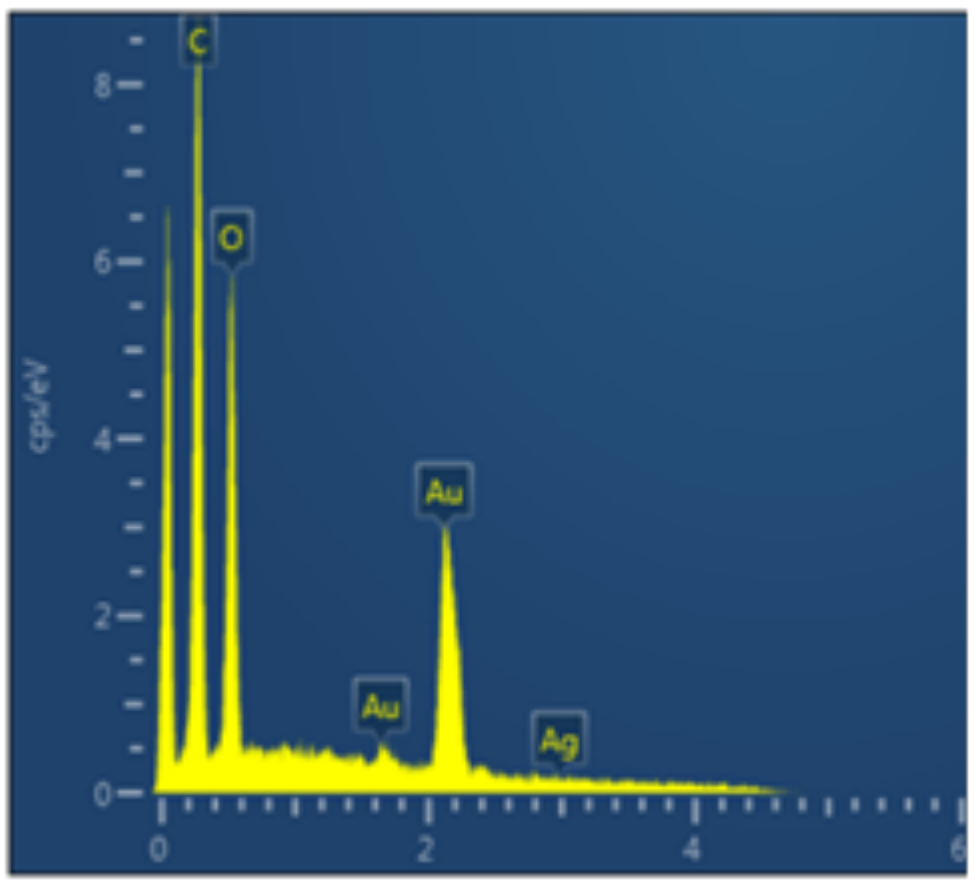

**Figure 6.** EDX analysis of AgMP034 without washing.

*Antibacterial Studies*

Cotton fabrics, as one of the most common types of textiles, are prone to the growth of microorganisms. Dangerous pathogens, particularly bacteria, can flourish in these fabrics and cause several problems [43]. Therefore, antibacterial textiles are essential in highly contaminated areas, including hospitals, as they can minimize the spread of pathogens. The main goal of this study was to develop highly durable textile fabrics with antimicrobial properties even after multiple washes. This was achieved by coating cotton fabrics with a gel-like silver-chitosan solution. To investigate the antibacterial properties of the coated fabric, AATCC 100 (quantitative analysis) and AATCC 147 (qualitative analysis) were utilized. Table 1 shows the killing efficiency of AgMP034 and SS26 after undergoing 75 wash cycles, and also A34 without washing and after 25, and 50 cycles following the AATCC 100 method. Untreated cotton fabric was also used as a control sample. According to Table 2, all treated samples, except A34, after 50 washing cycles, showed a 100 percent reduction for both *E. coli* and *S. aureus* even after 75 wash cycles. Despite decreased silver content due to washing, the remaining quantity surpassed the minimum Ag concentration necessary to sustain their antibacterial properties. Our preliminary studies indicated that 100% bacterial reduction on coated cotton fabrics required a minimum Ag coverage of around 7.5 μg/g of fabric.

To investigate the effect of the crosslinking agent on the durability of coated fabrics, sample A34 was prepared. Table 1 showed that after 50 wash cycles, the reduction percent decreased to 99.88 (Log 10 reduction of 2.927) and 99.81 (Log 10 reduction of 2.723) for *E. coli* and *S. aureus*, respectively. The lack of the crosslink agent causes a weaker interaction between silver particles and cotton fabrics.

**Table 1.** AATCC100- Effect of crosslinking and washing cycles on antibacterial properties

| Samples | Washing Cycles | *E. coli* | | *S. aureus* | |
|---|---|---|---|---|---|
| | | Percent Reduction | Log 10 Reduction | Percent Reduction | Log 10 Reduction |
| **Control** | 0 | 0 | 0 | 0 | 0 |
| **A34** | 0 | 100 | >6 | 100 | >6 |
| | 25 | 100 | >6 | 100 | >6 |
| | 50 | 99.88 | 2.927 | 99.81 | 2.723 |
| **AgMP034** | 0 | 100 | >6 | 100 | >6 |
| | 25 | 100 | >6 | 100 | >6 |
| | 50 | 100 | >6 | 100 | >6 |
| | 75 | 100 | >6 | 100 | >6 |
| **SS26** | 0 | 100 | >6 | 100 | >6 |
| | 25 | 100 | >6 | 100 | >6 |
| | 50 | 100 | >6 | 100 | >6 |
| | 75 | 100 | >6 | 100 | >6 |

The antibacterial properties of the uncoated and silver-coated cotton fabrics were also analyzed through the AATCC 147 method. Parallel streaks of both *S. aureus* as gram-positive and *E. coli* as gram-negative bacteria were drawn in separate Petri dishes and subsequently covered with the pieces of the samples. Figure 7 shows the photographs of the negative control and control inoculated with bacteria as a reference. The bacterial growth that occurred underneath the uncoated fabrics was visible. Also, there was no bacteria growth on agar for the negative control, indicating that the incubator's environment and samples were clean. Based on Figure 8, Sample A34 without washing and after 25 wash cycles showed no growth of bacteria, which indicated splendid antibacterial activities. However, after 50 wash cycles, there were bacteria colonies, which may be attributed to the lack of enough silver coverage on this sample. In fact, due to the lack of crosslinking agent, silver particles coated with chitosan lost their stability on the fabric and washed away during laundering. For sample A34 without washing, a slight inhibition zone was observed around the fabric, indicating the leaching of silver particles. It likely resulted from a weak interaction between the coating material and fabric, leading to gradual release by washing. However, there was no distinct boundary for sample A34 after 25 wash cycles.

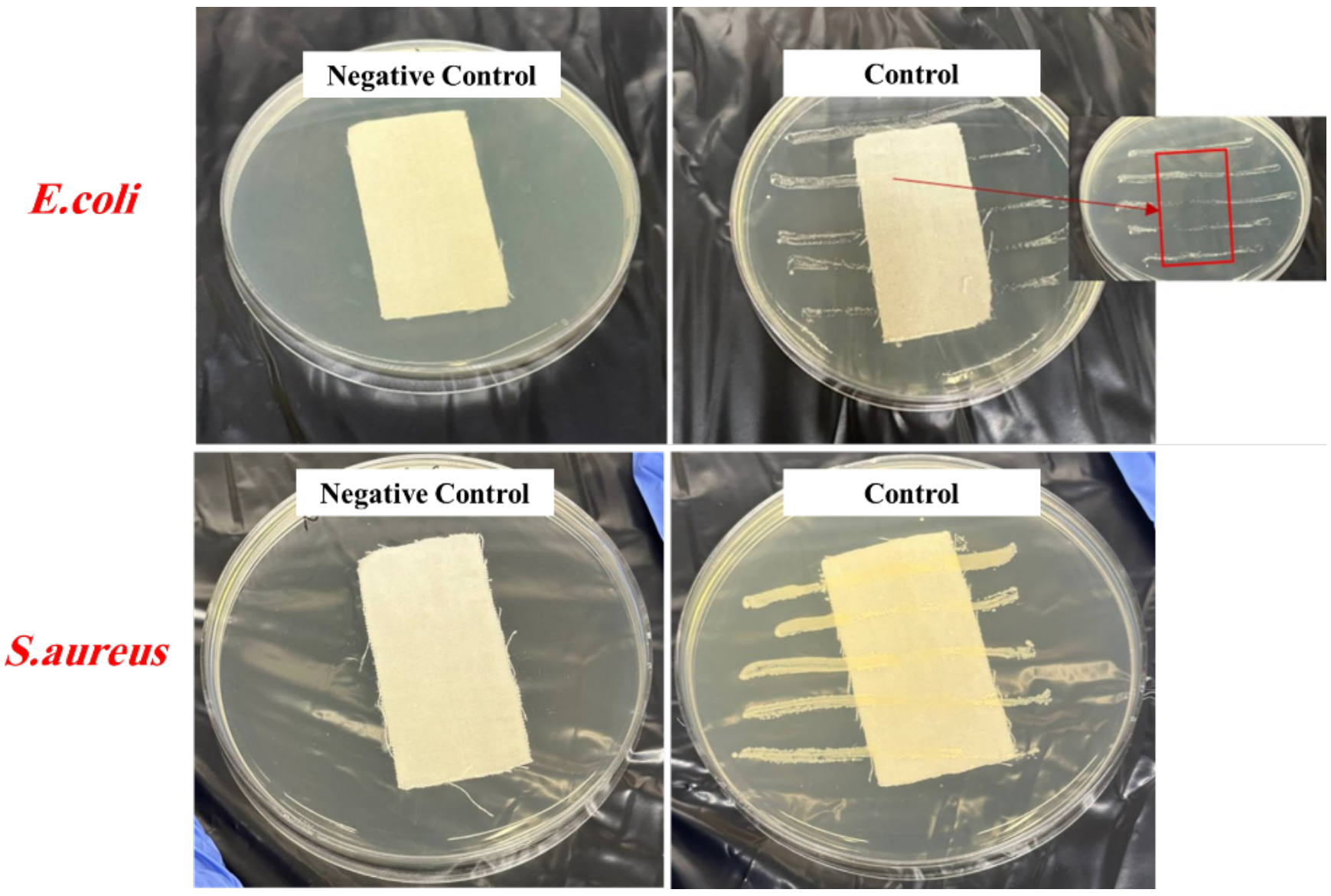

**Figure 7**. AATCC 147 for Negative control and untreated fabric against *E. coli and S. aureus*

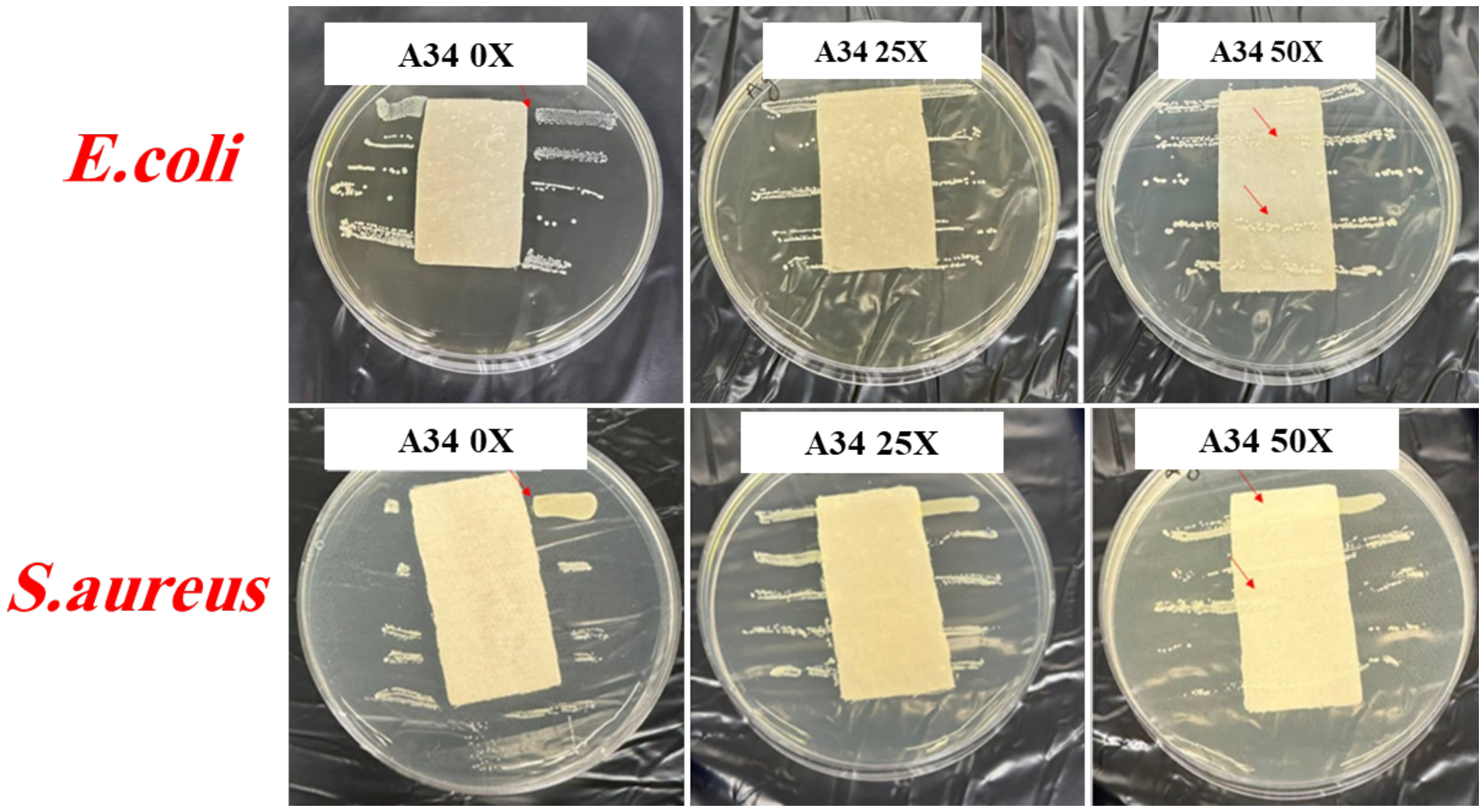

**Figure 8.** AATCC 147- A34 sample for No washing, 25, and 50 wash cycles against *E. coli and S. aureus*

To study the effect of the crosslinking agent on washing durability, the antibacterial activity of AgMP034 and SS26 was also tested qualitatively against the same bacterial strains used for previous samples, and results were shown in Figure 9 for *E. coli* and Figure 10 for *S. aureus*. Based on Figure 9 and Figure 10, all the coated samples showed remarkable antibacterial efficacy even after 75 wash cycles, which can be attributed to the sufficient Ag coverage on the samples. Therefore, highly durable coated fabrics can be obtained by using a crosslinking agent. Cross-linking agent helps to form a stable and durable coating on the cotton fabrics. It improves the adherence of the microparticles to the cotton fibers, resulting in sufficient antibacterial properties even after several washing cycles.

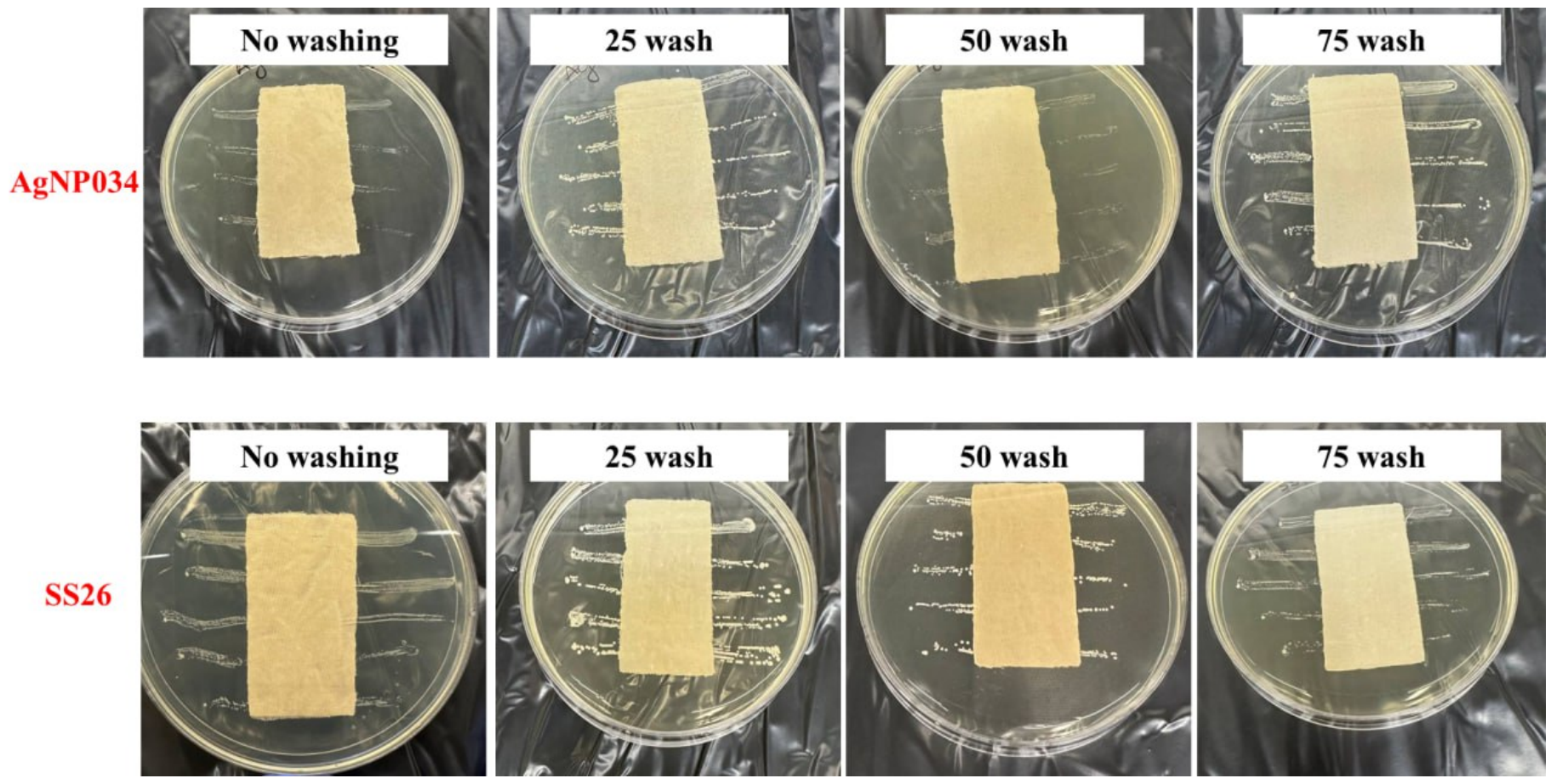

**Figure 9.** AATCC 147- AgMP034 and SS26 samples for No washing, 25, 50, and 75 wash cycles against *E. coli*

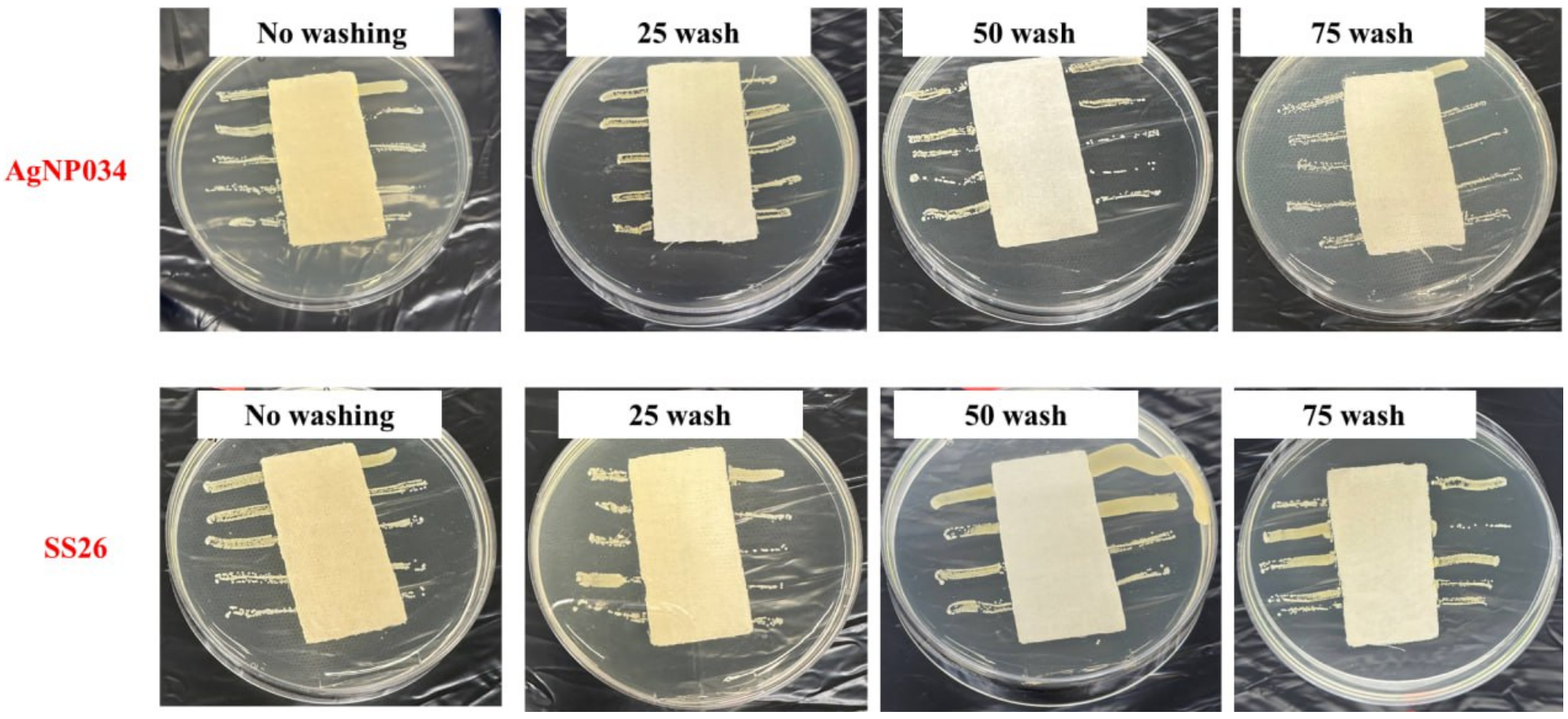

**Figure 10.** AATCC 147- AgMP034 and SS26 samples for No washing, 25, 50, and 75 wash cycles against *S. aureus*

*Bactericidal Action against Multidrug Resistant E. coli Strains*

We conducted pilot experiments comparing non-pathogenic *E. coli* MG1655 strain and extended-spectrum b-lactamase producing ESBL41 and ESBL146 *E. coli* isolates. Samples of unbleached muslin coated with the Silver Shell™ (A34 were tested according to the described above AATCC 100 protocol. According to Table 2, we have found strong (over 3 log) reduction for all tested strains regardless of using freshly coated textile samples or those that were washed 25 times. This confirmed prior tests with model *E. coli* and *S. aureus* strains (Table 1) and showed that in this setup Ag bactericidal action is similar towards antibiotic-sensitive and antibiotic-resistant ESBL isolates.

**Table 2.** AATCC100- Effect of coating and washing cycles on multidrug resistant *E. coli* Strains

| Samples | Washing Cycles | *E. coli MG1655* | | *ESBL* | |
|---|---|---|---|---|---|
| | | **Percent Reduction** | **Log 10 Reduction** | **Percent Reduction** | **Log 10 Reduction** |
| **Control** | 0 | 0 | 0 | 0 | 0 |
| **A34** | 0 | 99.971 | >3 | 99.99 | >4 |
| | 25 | 99.976 | >3 | 99.99 | >4 |

*Color Measurement Study*

The color measurements were carried out to study how the color of cotton fabrics is impacted by coating with gel-like Silver Shell™ solutions. These values are defined as L* for lightness, a* for red (+) to green (−), and b* for yellow (+) to blue (−) [9]. Tables 2 and 3 showed CIE L*a*b* values for SS26 and AgMP034 after 75 washing cycles, respectively. The color coordinates were also measured for untreated cotton for comparison. Based on the results, by applying the fabrics with Silver Shell™, L* was decreased and both a* and b* were increased. Although there was a slight change for a*, the significant change was attributed to b*, which showed higher values for treated samples. Therefore, the bluish appearance of untreated cotton would change to yellowness.

The quality of the coating was evaluated based on the CIEL*a*b* color difference formula as follows [44]:

$$\Delta E^*_{ab} = \sqrt{\Delta L^{*2} + \Delta a^{*2} + \Delta b^{*2}} \quad (3)$$

Measurements were randomly taken at five different locations and the color difference between each pair of these five spots was calculated, resulting in the average values presented in Table 2 and Table 3. Based on the results, the color difference between SS26 and AgMP034 was below 2. This suggests a slight color difference, indicating a uniformity in the coating. The average values of color difference for SS26 and AgMP034 before washing were 0.96 and 1.91, respectively.

**Table 2.** CIE L*a*b* values of SS26 samples

| Samples | L* | a* | b* | ΔE |
|---|---|---|---|---|
| Untreated Cotton | 95.73 ± 0.10 | 1.01 ± 0.19 | -3.95 ± 1.44 | 0.52 ± 0.23 |
| SS26- 0X | 79.27 ± 0.39 | 2.85 ± 0.34 | 5.21 ± 0.39 | 0.96 ± 0.38 |
| SS26- 25X | 87.63 ± 0.75 | 2.56 ± 0.36 | 7.29 ± 1.14 | 1.45 ± 0.56 |
| SS26- 50X | 90.24 ± 0.22 | 1.69 ± 0.05 | 5.43 ± 0.68 | 0.99 ± 0.55 |
| SS26- 75X | 92.52 ± 0.26 | 0.53 ± 0.13 | 4.83 ± 0.27 | 0.59 ± 0.21 |

**Table 3**. CIE L*a*b* values of AgMP034 samples

| Samples | L* | a* | b* | ΔE |
|---|---|---|---|---|
| Untreated Cotton | 95.73 ± 0.10 | 1.01 ± 0.19 | -3.95 ± 1.44 | 0.52 ± 0.23 |
| AgMP034- 0X | 82.24 ± 0.38 | 2.41 ± 0.18 | 9.97 ± 1.30 | 1.91 ± 1.01 |
| AgMP034- 25X | 88.71 ± 0.20 | 1.73 ± 0.29 | 5.59 ± 1.30 | 1.51 ± 0.71 |
| AgMP034- 50X | 90.54 ± 0.62 | 0.76 ± 0.05 | 5.94 ± 0.42 | 1.12 ± 0.42 |
| AgMP034- 75X | 93.13 ± 0.18 | 0.24 ± 0.07 | 3.89 ± 0.29 | 0.51 ± 0.20 |

Figure 11 illustrates the color strength (K/S) plots of SS26 and AgMP034 before and after washing. Based on Figure 11, the color strength decreased after washing for both samples, indicating a reduced silver amount on the fabrics. To elucidate the significance of the bond formation between microparticles and cotton fabric as a substrate, the %Fixation was compared for AgMP034 and SS26 following multiple wash cycles. According to Figure 12, SS26 exhibited %Fixation values of 44%, 32%, and 28% following 25, 50, and 75 washing cycles, respectively. For AgMP034, the % fixation values were higher at 54%, 44%, and 32% for the same sequence of washing cycles. The presence of chitosan surrounding the silver particles might be the reason for the higher % fixation value for AgMP034.

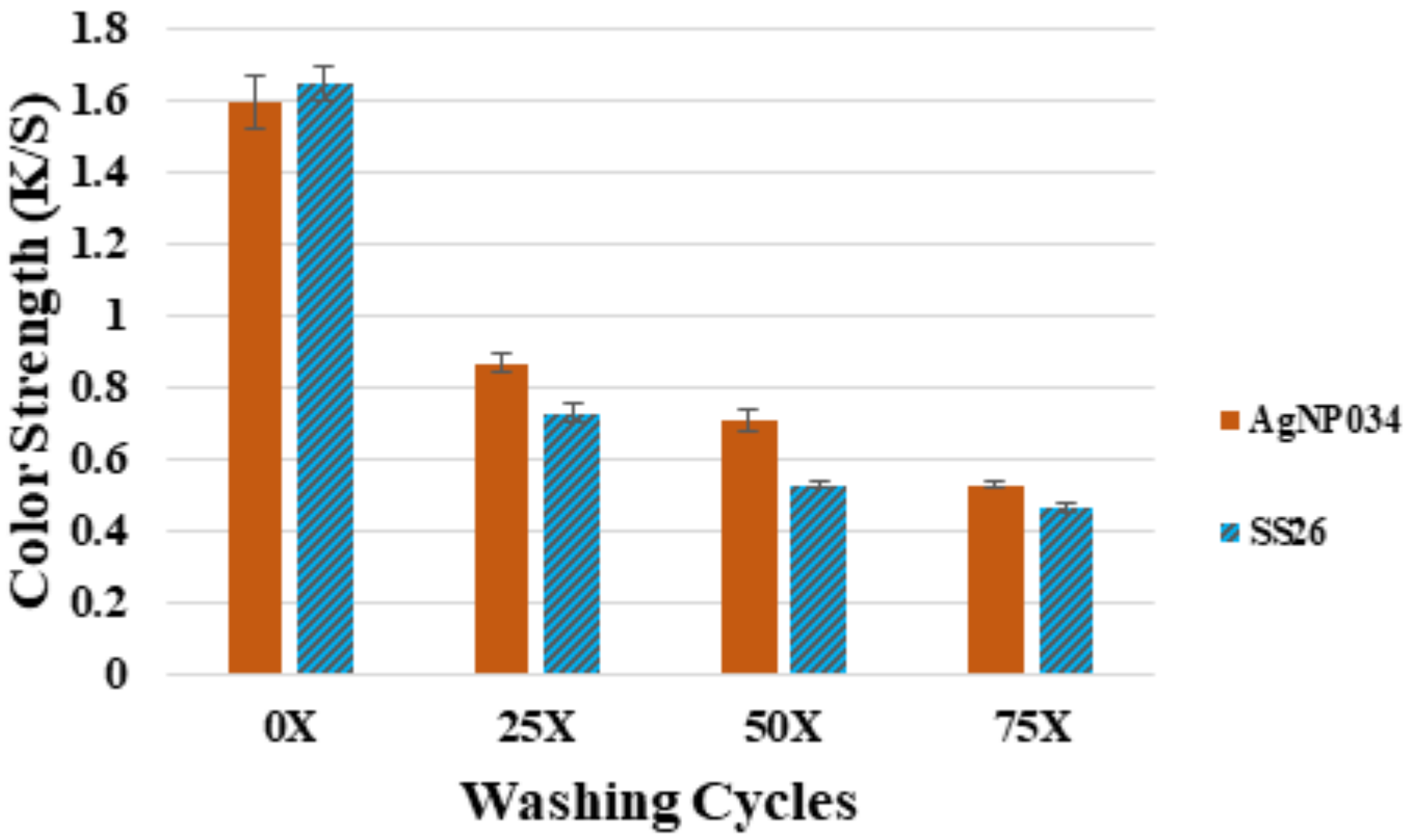

**Figure 11**. Color Strength of AgMP034 and SS26 samples in different washing conditions

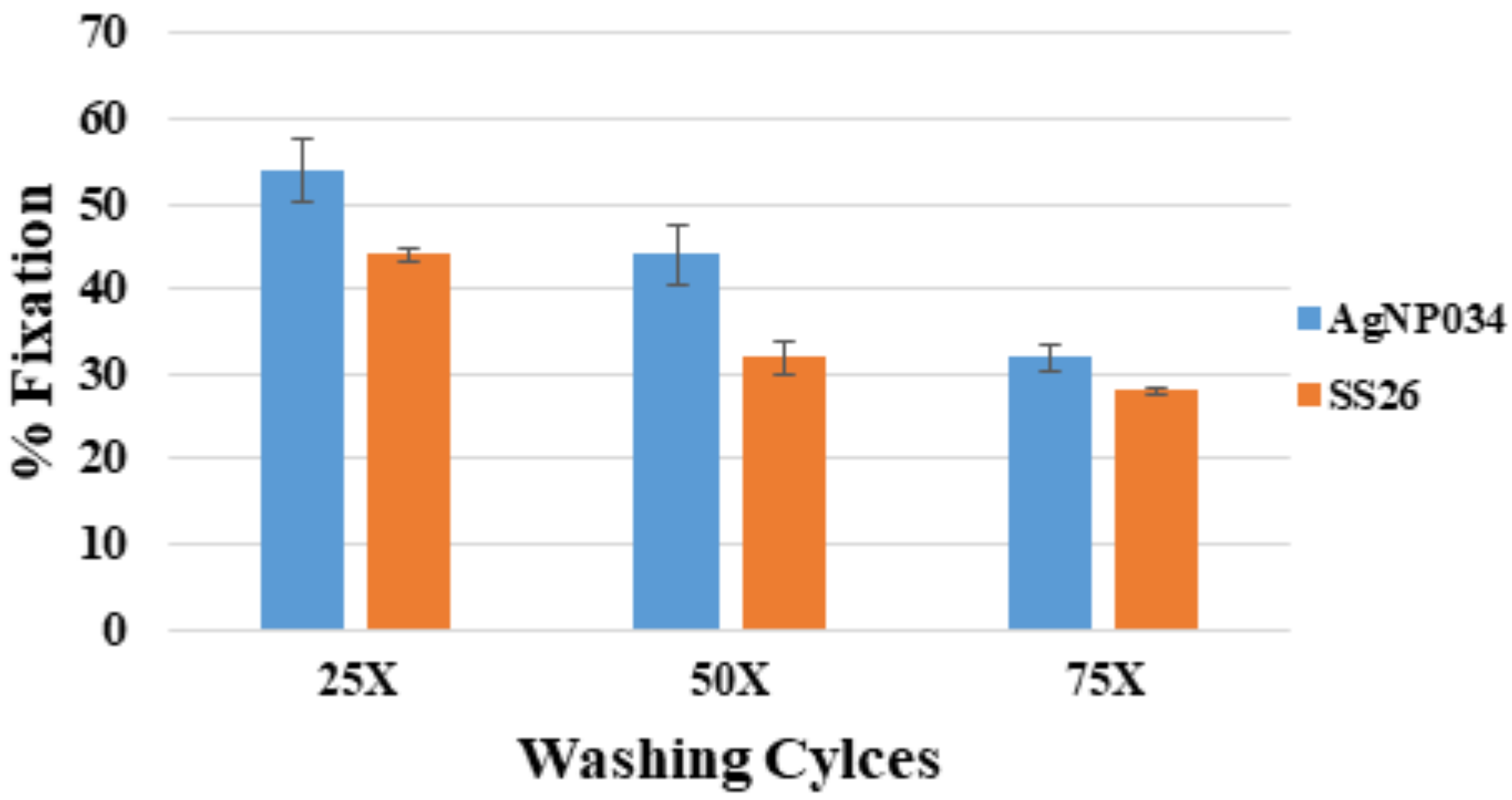

**Figure 12.** Fixation (%F) of AgMP034 and SS26 samples in different washing conditions

## 4. Conclusions

In summary, the present study provided data on structures and antibacterial efficiency of coated cotton fabrics using Silver Shell™ solution. The results indicated the valuable and innovative Silver Shell™ Solution to combat dangerous hospital-acquired infections and ensured long-term antimicrobial activity in textiles. According to TEM results, Silver Shell™ Solution has a core-shell structure, and with chitosan forms a protective shell around silver microparticles. The coating quality of fabrics was investigated by FESEM before and after multiple wash cycles, indicating highly durable washing fastness. The AATCC-100 antibacterial results indicated a 100% reduction for two types of bacteria, *S. aureus* and *E. coli*, even after 75 wash cycles. The antibacterial results for the coated samples without using a crosslinking agent exhibited a 99.88% and 99.81% reduction for

*S. aureus* and *E. coli*, respectively after 50 wash cycles. To investigate the leaching properties of coated samples with Silver Shell™ solution and study the effect of the crosslinking agent, AATCC-147 was performed. Despite the remarkable antibacterial efficacy of coated samples even after 75 washing cycles, the crosslinking agent minimized the release of submicron particles from the fabric. According to the results of color measurements, uniform coatings were obtained in different samples. Fixation values of SS26 samples showed 44%, 32%, and 28% following 25, 50, and 75 washing cycles, respectively. For AgMP034, the % fixation values were higher than SS26 and calculated at 54%, 44%, and 32% for the same sequence of washing cycles. The presence of chitosan surrounding the silver particles might be the reason for the higher fixation value for AgMP034.

**Author Contributions:** Conceptualization, S.M., D.J. and V.R.; methodology, S.M. and V.R.; software, S.M.; validation, V.R., T.A.S. and D.A.V.; formal analysis, S.M., D.J. and V.K.; investigation, S.M., D.J., K.P., A.S., D.P. and V.K.; resources, D.A.V., V.R. and M.P.; data curation, V.R. and K.P.; writing—original draft preparation, S.M.; writing—review and editing, V.R., K.P. and T.A.S.; visualization, S.M., D.J., A.S., D.P., T.A.S. and V.K.; supervision, V.R. and D.A.V.; project administration, V.R. and D.A.V.; funding acquisition, D.A.V. All authors have read and agreed to the published version of the manuscript.

**Funding:** Research reported in this publication was supported by the National Institute of Allergy and Infectious Diseases of the National Institutes of Health under Award Number R41AI172693. The content is solely the responsibility of the authors and does not necessarily represent the official views of the National Institutes of Health.

**Data Availability Statement:** All data is available within the manuscript.

**Conflicts of Interest:** The authors declare no conflict of interest.